\documentclass{article}
\usepackage{graphicx}
\usepackage{cite}
\usepackage{amsmath}

\begin{document}

\title{Scheduling Wireless Links in the Physical Interference Model by
Fractional Edge Coloring}

\author{Guilherme I. Ricardo\\
Jos\'e F. de Rezende\\
Valmir~C.~Barbosa\thanks{Corresponding author (valmir@cos.ufrj.br).}\\
\\
Programa de Engenharia de Sistemas e Computa\c c\~ao, COPPE\\
Universidade Federal do Rio de Janeiro\\
Caixa Postal 68511\\
21941-972 Rio de Janeiro - RJ, Brazil}

\date{}

\maketitle

\begin{abstract}
We consider the problem of scheduling the links of wireless mesh networks for
capacity maximization in the physical interference model. We represent such a
network by an undirected graph $G$, with vertices standing for network nodes and
edges for links. We define network capacity to be $1/\chi'^*_\mathrm{phys}(G)$,
where $\chi'^*_\mathrm{phys}(G)$ is a novel edge-chromatic indicator of $G$,
one that modifies the notion of $G$'s fractional chromatic index. This index
asks that the edges of $G$ be covered by matchings in a certain optimal way. The
new indicator does the same, but requires additionally that the matchings used
be all feasible in the sense of the physical interference model. Sometimes the
resulting optimal covering of $G$'s edge set by feasible matchings is simply a
partition of the edge set. In such cases, the index $\chi'^*_\mathrm{phys}(G)$
becomes the particular case that we denote by $\chi'_\mathrm{phys}(G)$, a
similar modification of $G$'s well-known chromatic index. We formulate the exact
computation of $\chi'^*_\mathrm{phys}(G)$ as a linear programming problem, which
we solve for an extensive collection of random geometric graphs used to
instantiate networks in the physical interference model. We have found that,
depending on node density (number of nodes per unit deployment area), often $G$
is such that $\chi'^*_\mathrm{phys}(G)<\chi'_\mathrm{phys}(G)$. This bespeaks
the possibility of increased network capacity by virtue of simply defining it so
that edges are colored in the fractional, rather than the integer, sense.

\bigskip
\noindent
\textbf{Keywords:}
Wireless mesh networks,
Link scheduling,
Physical interference model,
Edge coloring,
Fractional edge coloring.
\end{abstract}

\newpage
\section{Introduction}
\label{intr}

We consider a set of nodes operating under the constraints imposed by the
physical interference model of wireless communication \cite{Gupta2000}. These
nodes are interconnected by a set $L$ of links, each link $e\in L$ being
characterized by a sender node $s_e$ and a receiver node $r_e$. Any node may in
principle act either as sender or as receiver, depending on the links in which
it participates. When all links in a set $S\subseteq L$ are concomitantly
active, the ability of receiver $r_e$ to decode what it receives from sender
$s_e$ for any given link $e\in S$ is constrained by the
signal-to-interference-plus-noise ratio (SINR) that results from the combined
activity of the group, given by \begin{equation}
\mathrm{SINR}(e,S)=\frac
{P/d_{s_er_e}^\alpha}
{\gamma+\sum_{f\in S\setminus\{e\}}P/d_{s_fr_e}^\alpha}.
\label{sinr}
\end{equation}
In this expression, $P$ is a node's transmission power (assumed the same for all
nodes), $\gamma$ is the noise floor, $d_{ab}$ is the Euclidean distance between
nodes $a$ and $b$, and $\alpha>2$ is used to determine how power decays away
from the transmitter with the distance to it.

The way the SINR constraint operates is by affecting the so-called feasibility
of set $S$. Specifically, we say that a nonempty $S\subseteq L$ is
\emph{feasible} if no two of its links share a node and, additionally,
$\mathrm{SINR}(e,S)\ge\beta$ for all $e\in S$, where $\beta>1$ is a parameter
related to a receiver's decoding capabilities, assumed the same for all
receivers. For consistency, we assume that the link set $L$ is such that, for
any singleton $\{e\}\subseteq L$, it holds that
$P/\gamma d_{s_er_e}^\alpha\ge\beta$. That is, every link in $L$ is capable of
providing effective communication from its sender to its receiver when operating
in isolation. Equivalently, set $\{e\}$ is assumed feasible.

The problem of maximizing network capacity, broadly understood as the rate of
effective communication among nodes, is closely related to that of scheduling
the links in $L$ for operation. This, in turn, is often posed under the
so-called physical interference model (in which the SINR constraint is fully
taken into account) but sometimes assumes only the constraints imposed by the
so-called protocol-based interference model (which depend essentially on
graph-based distances). Solving the link-scheduling problem may require
computationally hard combinatorial problems to be tackled and has given rise to
numerous proposals, some approaching the scheduling problem by itself
\cite{Jain2003,
Brar2006,
Blough2010,
Wang2006,
Moscibroda2007,
Chafekar2008,
Hua2008,
Goussevskaia2009,
Santi2009,
Yang2009,
Boyaci2010,
Yang2010,
Augusto2011,
Leconte2011,
Shi2011,
Goussevskaia2014,
Nabli2014,
Wang2015,
Zhou2015,
vrb16},
some in conjunction with others
\cite{hs88,
Cruz2003,
Alicherry2005,
Wang2008,
Capone2010,
Gore2011,
Kesselheim2011,
Wan2011,
Rubin2012,
Vieira2012}.

Many of these proposals are formulated within the framework of spatial
time-division multiple access (STDMA), which divides time into slots and reduces
the scheduling problem to the selection of which links to activate
simultaneously in each one. The type of selection strategy that is by far the
most adopted asks that a sequence
$\mathcal{S}=\langle S_1,S_2,\ldots,S_T\rangle$ be determined for some $T>1$. In
this sequence, each $S_i$ is a subset of the link set $L$ complying with the
constraints imposed by the interference model in use and moreover ensuring that
each $e\in L$ appears in exactly one of the $T$ subsets. Any approach to
maximize network capacity by solving the link-scheduling problem will therefore
seek to minimize $T$ (maximize $1/T$) without violating any of these
constraints. Once a solution is available, repeating the sequence $\mathcal{S}$
guarantees interference-free communication for as long as needed. 

One common abstraction for reasoning about such proposals is that of graph
coloring or related notions. Of the proposals mentioned above, some do indeed
make explicit reference to such an abstraction
\cite{hs88,
Jain2003,
Hua2008,
Yang2009,
Vieira2012,
Nabli2014,
vrb16}.
In terms of the formulation outlined above, clearly the links in each set $S_i$,
to be scheduled for operation in the same time slot, can be regarded as being
assigned the same color (the $i$th of the $T$ available colors) if we interpret
the conditions for membership in $S_i$ in the context of graph coloring. Whether
it is vertex coloring or edge coloring that is being considered depends on how
the graph in question is set up to represent how the various links relate to
each other given the interference model at hand. In either case, once the least
possible value of $T$ is found (or approximated), each vertex or edge ends up
having exactly one color.

To the best of our knowledge, the decades-long effort to come up with
capacity-optimizing strategies for link scheduling has almost completely failed
to recognize that such a single-color abstraction is inherently limited and may
in many cases fall short of leading to as much network capacity as possible. The
exceptions we know of are only three and separated by many years. The earliest
one is based on the coloring of a graph's edges and adopts what would pass for
the protocol-based interference model had it existed at the time \cite{hs88}.
The other two are very recent and both based on the coloring of vertices, the
earliest one given for the protocol-based interference model as well
\cite{Vieira2012}, the latest for the physical interference model \cite{vrb16}.

What these three proposals have in common is that they use the fractional
variety of graph coloring. In terms of the STDMA scheme explained above, what
they do is let each link appear in exactly $q\ge 1$ of the $T$ sets in
$\mathcal{S}$, the same value of $q$ for all links, instead of in one single
set. Moreover, instead of minimizing $T$ alone, they seek to minimize the ratio
$T/q$ while treating $q$ as a variable. If $T^1$ is the least number of slots to
accommodate all $\vert L\vert$ link activations, one per link, in the
single-color case, and if the pair $(T^*,q^*)$ provides the least possible
$T/q$ ratio while accommodating all $q\vert L\vert$ link activations, $q$ per
link, in the fractional-coloring case, then conceivably it may happen that
$T^*/q^*<T^1$. If this does happen, then clearly we have $T^*<T^1q^*$, so the
$T^*$-slot sequence is shorter than $q^*$ repetitions of the $T^1$-slot sequence
and therefore the former is preferable to the latter, since in the two cases we
have the same total number of link activations, viz., $q^*\vert L\vert$.
Readily, in this case the sequence to be repeated in order for
interference-free communication to be provided while needed is the one
comprising $T^*$ sets. Thus, so far as we seek to maximize network capacity via
link scheduling, what needs to be done is maximize the ratio $q/T$. This ratio
is how we define network capacity henceforth.

Important though the fractional-coloring-based contributions in
\cite{hs88,Vieira2012,vrb16} have been, they have each left relevant problems
open as well. As we see these problems, the most relevant one is the search for
approaches for the exact determination of optimal fractional colorings in the
physical interference model. As we remarked above, this has been attempted
neither by the proposals in \cite{hs88,Vieira2012} (both of which target the
protocol-based interference model, though the former is exact while the latter
is a heuristic) nor by the one in \cite{vrb16} (which is a heuristic, even if
one for the physical interference model).

Our aim in this paper is to make some headway toward achieving such an exact
method. As will become clear along the way, the difficulty lies not so much in
the possibility of obtaining such an exact method of solution, which we do
describe in detail, but rather in making its computational hardness scale in
such a way that tackling large instances is still possible. The exact method we
describe has allowed us to chart the landscape of a class of random networks
regarding the possibility of fractional-coloring-based link scheduling that is
superior to its single-color counterpart. We have been able to do this for a
reasonably wide range of the parameters involved in network generation, which
has led us to conclude that, given the uncertainties afforded by the confidence
intervals obtained, networks with the potential to benefit from a
fractional-coloring approach occur in a non-negligible proportion. These
empirical findings, along with the implicit impetus they lend to the search for
approaches that are more computationally efficient, constitute one of our
contributions. (We also pinpoint possible alternatives that might lead to more
scalable approaches, but finding them remains an open issue.) Our main
contribution, though, is that by formulating capacity optimization in the
framework of fractional coloring, we automatically provide for the fallback
solution in which it is integer (single-color) coloring, rather than nontrivial
fractional (more-than-one-color) coloring, that yields optimal capacity. Integer
coloring, after all, is simply the particular case of fractional coloring in
which it is better to use one, rather than more than one, color per link.

Before we proceed, a few remarks are in order about the particular view of
network-capacity maximization we have adopted, that is, maximization via link
scheduling in the physical interference model. While this view is the same as in
\cite{Gupta2000,Brar2006,Goussevskaia2009,Goussevskaia2014,vrb16}, there are
nevertheless various other issues that are sometimes taken into account. These
include node placement \cite{Mathar2000,Prommak2002,Wertz2004,Bahri2005},
frequency assignment \cite{Narayanan2002,Aardal2007}, as well as taking
end-to-end communication demands into consideration
\cite{Mathar2000,Narayanan2002,Prommak2002,Wertz2004,Bahri2005} or not
\cite{Aardal2007}. Moreover, even though some approaches do resort to vertex
coloring that assigns more than one color to the same vertex
\cite{Narayanan2002,Brar2006}, they have remained oblivious to the fact that the
real power of coloring the same vertex or edge by multiple colors lies in the
potential to exploit the graph's fractional-coloring properties (i.e., through
the minimization of a rational, not an integer, number), not in color
multiplicity per se.

We continue in the following manner. First, in Section~\ref{math}, we give a
detailed mathematical formulation of the fractional-coloring-based approach to
link scheduling that we pursue. Then we move to Section~\ref{setup}, where our
computational methodology is laid out. Our results are presented in
Section~\ref{results}, with discussion, and we finalize in Section~\ref{concl}
with a summary, concluding remarks, and comments on future prospects.

\section{Mathematical formulation}
\label{math}

Given the set $L$ of links to be scheduled, and letting $N$ be the set of all
nodes acting as sender or receiver in at least one link in $L$, we consider the
undirected graph $G=(N,L)$, that is, the graph having $N$ for set of vertices
and $L$ for set of undirected edges. Our use of a graph and not a multigraph
(which would allow multiple edges joining the same two vertices) is meant to
allow us to use the same experimental setting as in \cite{vrb16}. As we discuss
in Section~\ref{setup}, in that setting no two links are allowed to interconnect
the same two nodes, not even in different directions (i.e., with their roles as
sender or receiver reversed). We do make this assumption about $L$ henceforth,
but the reader is to note that no further modeling difficulties would arise
otherwise (cf.\ Section~\ref{multi}).

We begin with the presentation of a linear programming (LP) problem for the
determination of maximum network capacity as defined in Section~\ref{intr}. That
is, we aim to formulate the problem of finding the integers $T$ and $q$ that
minimize the ratio $T/q$ while allowing every link in $L$ to be active in
exactly $q$ of $T$ time slots while respecting the constraints imposed by the
physical interference model.

\subsection{LP problem}
\label{LP}

By the definition of a feasible set of links and also the definition of graph
$G$ above, clearly every feasible set of links corresponds to a matching in $G$,
though the converse may not be true. Henceforth we refer as a \emph{feasible
matching} to any matching whose edges constitute a feasible set of links.

Let $\mathcal{M}$ be the set of all feasible matchings of $G$. For each
$M\in\mathcal{M}$, let $x_M$ be a real variable and consider the following LP problem.
\begin{align}
\text{minimize } &\textstyle w=\sum_{M\in\mathcal{M}}x_M & \label{xobj}\\
\text{subject to }
&x_M\ge 0, &\forall M\in\mathcal{M} \label{xnonneg}\\
&\textstyle\sum_{M\in\mathcal{M}\vert e\in M}x_M=1.
&\forall e\in L \label{xsumis1}
\end{align}
This problem asks that the sum of all $x_M$'s (the objective function $w$ in
Eq.~(\ref{xobj})) be minimized while respecting the constraints that none of
them be allowed to become negative (Eq.~(\ref{xnonneg})) and that, for each edge
$e\in L$, those $x_M$'s for which $e\in M$ add up to $1$ (Eq.~(\ref{xsumis1})).
Because the coefficients of the $x_M$'s in Eqs.~(\ref{xobj}) and~(\ref{xsumis1})
are all equal to $1$, hence rational numbers, at least one solution exists
minimizing $w$ with every $x_M$ a rational number as well. Let $\mathcal{P}$ be
the subset of $\mathcal{M}$ such that $M\in\mathcal{P}$ if and only if $x_M>0$
in this solution.

For $M\in\mathcal{P}$, let $p_M/q_M$ be such positive rational value of $x_M$
minimizing $w$. If $q^*$ denotes the least common multiple of all $q_M$'s over
$M\in\mathcal{P}$, then the desired minimum value of $w$, call it $w^*$, can be
written as
\begin{equation}
w^*=\frac{\sum_{M\in\mathcal{P}}T_M}{q^*},
\label{optm}
\end{equation}
where $T_M=q^*p_M/q_M$ is necessarily a positive integer.

Now consider any edge $e\in L$ and let $\mathcal{P}_e$ be the subset of
$\mathcal{P}$ such that $M\in\mathcal{P}_e$ if and only if $e\in M$. That is,
$\mathcal{P}_e$ is the set of all feasible matchings $M$ of $G$ that contribute
to the minimum value of $w$ with a positive $x_M$ and moreover include edge $e$.
Set $\mathcal{P}_e$ is necessarily nonempty, since the matching containing $e$
and no other edge is by assumption feasible. By the constraint in
Eq.~(\ref{xsumis1}), we have
\begin{equation}
\sum_{M\in\mathcal{P}_e}T_M=q^*\sum_{M\in\mathcal{P}_e}\frac{p_M}{q_M}=q^*.
\end{equation}
If we view each $T_M\ge 1$ as a sort of multiplicity of matching $M$, then this
equation is saying that the added multiplicities of all matchings in
$\mathcal{P}_e$ equals $q^*$.

What this means in the context of scheduling the links in $L$ is that, if we let
all links in $M$ be concomitantly active for $T_M$ time slots and do this for
all $M\in\mathcal{P}_e$, then after all $\sum_{M\in\mathcal{P}_e}T_M$ time slots
link $e$ will have appeared $q^*$ times, regardless of the particular link $e$
under consideration. Thus, ensuring that this happens for every $e\in L$
requires
\begin{equation}
\sum_{M\in\cup_{e\in L}\mathcal{P}_e}T_M=\sum_{M\in\mathcal{P}}T_M
\end{equation}
time slots. We denote this overall number of time slots by $T^*$ and, by
Eq.~(\ref{optm}), conclude that $w^*=T^*/q^*$ is the desired minimum value of
the ratio $T/q$.

\subsection{Edge-coloring interpretation}
\label{ecoloring}

If we allow $\mathcal{M}$ to include every one of the graph's matchings, without
regard to how any particular matching stands as far as our link-scheduling
problem is concerned, then in graph-theoretic terms the preceding development
explains why the LP problem given in Eqs.~(\ref{xobj})--(\ref{xsumis1}) can be
taken as defining the fractional chromatic index of $G$
\cite{sstf12}.\footnote{Alternatively, multichromatic index \cite{fs77},
fractional edge chromatic number \cite{s79,su97,bm08}, or fractional
edge-coloring number \cite{s03} of $G$.} This index, which we denote by
$\chi'^*(G)$, reflects the most ``efficient'' way in which the graph's set of
edges can be covered by $T$ matchings in such a way as to let each edge belong
to exactly $q$ of the matchings. The use of efficient here refers to the
minimization of the ratio $T/q$, hence the graph-theoretic interpretation in the
case of an all-encompassing $\mathcal{M}$. The well-known, alternative
definition of $\chi'^*(G)$ as
\begin{equation}
\chi'^*(G)=\min_{k\ge 1}\frac{\chi'^k(G)}{k}
\label{chi*}
\end{equation}
is an easy consequence of the same development. In this expression, $\chi'^k(G)$
is the minimum number of matchings needed to cover $G$ in such a way that every
edge belongs to exactly $k$ matchings. Setting $k=1$ yields the usual chromatic
index of $G$, $\chi'(G)=\chi'^1(G)$, for which it then holds that
\begin{equation}
\chi'^*(G)\le\chi'(G).
\label{chi*b}
\end{equation}

By analogy, restricting $\mathcal{M}$ to include only feasible matchings admits
a graph-theoretic interpretation as well. In this interpretation, the edge set
$L$ of $G$ has to be covered by $T$ feasible matchings while mandatorily
including every edge in exactly $q$ of them and minimizing $T/q$.
A problem-specific fractional chromatic index can then be defined for $G$ based
on Eqs.~(\ref{xobj})--(\ref{xsumis1}), one that takes into account all the
specificities of the physical interference model discussed in
Section~\ref{intr} by requiring all members of $\mathcal{M}$ to be feasible. We
denote this new index by $\chi'^*_\mathrm{phys}(G)$ and generalize
Eqs.~(\ref{chi*}) and~(\ref{chi*b}) in the obvious way, obtaining
\begin{equation}
\chi'^*_\mathrm{phys}(G)=\min_{k\ge 1}\frac{\chi'^k_\mathrm{phys}(G)}{k}
\label{chi*phys}
\end{equation}
and
\begin{equation}
\chi'^*_\mathrm{phys}(G)\le\chi'_\mathrm{phys}(G),
\label{ineq}
\end{equation}
where each $\chi'^k_\mathrm{phys}(G)$ is defined analogously to $\chi'^k(G)$ and
$\chi'_\mathrm{phys}(G)=\chi'^1_\mathrm{phys}(G)$. In Section~\ref{concl}, we
discuss how computationally hard it may be to determine
$\chi'^*_\mathrm{phys}(G)$, especially vis-\`a-vis the determination of
$\chi'^*(G)$. Be that as it may, clearly network capacity as defined in
Section~\ref{intr} is given by $1/\chi'^*_\mathrm{phys}(G)$.

\subsection{Finding out whether
\boldmath $\chi'^*_\mathrm{phys}(G)<\chi'_\mathrm{phys}(G)$}
\label{steps}

One of the core elements of our study in this paper is the determination, for
some given $G$, of whether coloring its edges fractionally is more efficient (in
the sense explained earlier) than coloring them with one single color per edge.
Put differently, for each $G$ we must be able to determine whether the
inequality in Eq.~(\ref{ineq}) is strict, which clearly is true if and only if
the most efficient coloring of $G$'s edges employs $k>1$ colors per edge. Recall
that solving the LP problem in Eqs.~(\ref{xobj})--(\ref{xsumis1}) already gives
us the value of $\chi'^*_\mathrm{phys}(G)$ along with the corresponding $x_M$'s
that are nonzero. It would then seem that checking whether all of these $x_M$'s
equal $1$ suffices, since if they do we can immediately conclude that
$\chi'^*_\mathrm{phys}(G)=\chi'_\mathrm{phys}(G)$. However, that LP problem may
admit several optimal solutions, including some that involve non-unit $x_M$'s
even when another equally optimal solution involves unit $x_M$'s only. For this
reason, testing whether every nonzero $x_M$ equals $1$ in the optimal solution
returned by the LP solver is meaningful only in the affirmative case. In the
negative case the test is meaningless, since it does not necessarily follow that
$\chi'^*_\mathrm{phys}(G)<\chi'_\mathrm{phys}(G)$ (cf.\ Section~\ref{examples}
for an example).

Given this difficulty, our approach is to address the direct calculation of
$\chi'_\mathrm{phys}(G)$ as well. We do this by modifying the LP program of
Eqs.~(\ref{xobj})--(\ref{xsumis1}) so that each $x_M$ must be an integer equal
to $0$ or $1$. The result is the following integer linear programming (ILP)
problem.
\begin{align}
\text{minimize } &\textstyle w_\mathrm{int}=
\sum_{M\in\mathcal{M}}x_M & \label{xobjint}\\
\text{subject to }
&x_M\in\{0,1\}, &\forall M\in\mathcal{M} \label{x0or1}\\
&\textstyle\sum_{M\in\mathcal{M}\vert e\in M}x_M=1.
&\forall e\in L \label{xsumis1int}
\end{align}
Clearly, any valuation of the $x_M$'s satisfying the constraints in
Eqs.~(\ref{x0or1}) and~(\ref{xsumis1int}) characterizes a partition of the link
set $L$ into feasible matchings (specifically, a matching $M\in\mathcal{M}$ is
in the partition if and only if $x_M=1$). The objective function in
Eq.~(\ref{xobjint}) counts the corresponding number of matchings and therefore
its optimal value, call it $w^*_\mathrm{int}$, is such that
$w^*_\mathrm{int}=\chi'_\mathrm{phys}(G)$.

In summary, the following is how we find out out whether
$\chi'^*_\mathrm{phys}(G)<\chi'_\mathrm{phys}(G)$.
\begin{enumerate}
\item Find $\chi'^*_\mathrm{phys}(G)$ by solving the LP problem in
Eqs.~(\ref{xobj})--(\ref{xsumis1}).
\item If every nonzero $x_M$ in the solution equals $1$, then conclude that
$\chi'^*_\mathrm{phys}(G)=\chi'_\mathrm{phys}(G)$ and stop.
\item Find $\chi'_\mathrm{phys}(G)$ by solving the ILP problem in
Eqs.~(\ref{xobjint})--(\ref{xsumis1int}).
\item Test whether $\chi'^*_\mathrm{phys}(G)<\chi'_\mathrm{phys}(G)$.
\end{enumerate}
Clearly, Steps~1--4 amount to solving the ILP problem only in those cases in
which the solution to the LP problem is inconclusive as far as comparing
$\chi'^*_\mathrm{phys}(G)$ and $\chi'_\mathrm{phys}(G)$ is concerned.

\section{Experimental setup}
\label{setup}

Given a fixed graph $G=(N,L)$, of vertex set $N$ and edge set $L$, the
computational core of our experiments is carrying out Steps~1 and~3 of
Section~\ref{steps}, which solve an LP problem and an ILP problem on $G$,
respectively. In our experiments, graph $G$ is an instance of the following
random geometric graph. Given a $d\times d$ region in two-dimensional Euclidean
space, each of the $\vert N\vert$ vertices is placed in it uniformly at random.
For vertices $a,b\in N$, the unordered pair $(a,b)$ is an edge of $L$ if and
only if $d_{ab}\le (P/\beta\gamma)^{1/\alpha}$ (equivalently, if and only if
$\mathrm{SINR}((a,b),\{(a,b)\})\ge\beta$, where the role taken up by $a$ as
sender or receiver relative to $b$ is immaterial). Put differently, $(a,b)$ is
an edge in $G$ if and only if the singleton $\{(a,b)\}$ is feasible. For each
$e=(a,b)\in L$, sender $s_e$ is either $a$ or $b$ uniformly at random, with
receiver $r_e$ set correspondingly. This random graph is equivalent to what is
called a type-I network in \cite{vrb16}. We use $\alpha=4$, $\beta=316.23$ ($25$
dB), and $\gamma=8\times 10^{-11}$ mW ($-100.97$ dBm), as well as $P=300$ mW
($24.78$ dBm) throughout all experiments.

For fixed $\vert N\vert$ and $d$, we generated $1\,000$ graph instances and
tested each one for suitability to Steps~1--4. Failing instances were dropped,
so all our results are expressed as averages over the passing instances. An
instance can fail for at least one of three reasons: edge set $L$ is empty; edge
set $L$ has more than $128$ edges, in which case we lack the computational
resources to enumerate all feasible matchings that go in set $\mathcal{M}$; the
number of feasible matchings in set $\mathcal{M}$ is greater than
$50\times 10^6$, which is as far as we can go given $128$ GB of RAM and given
that we use the Gurobi suite (www.gurobi.com) for solving both the LP and ILP
problems, always with the pre-solver disabled and Simplex as the core linear
programming solver (we found that these Gurobi settings require the least amount
of RAM overall). The number of passing instances for each combination of
$\vert N\vert$ and $d$ values we used is given in Table~\ref{table1}.

\begin{table}
\centering
\caption{Number of graph instances used for each $\vert N\vert,d$ combination.}
\label{table1}
\small
\begin{tabular}{rrrrrrrrrrr}
\hline
&\multicolumn{10}{c}{$d$ (km)} \\
\cline{2-11}
$\vert N\vert$
&\multicolumn{1}{c}{$1$}
&\multicolumn{1}{c}{$2$}
&\multicolumn{1}{c}{$3$}
&\multicolumn{1}{c}{$4$}
&\multicolumn{1}{c}{$5$}
&\multicolumn{1}{c}{$6$}
&\multicolumn{1}{c}{$7$}
&\multicolumn{1}{c}{$8$}
&\multicolumn{1}{c}{$9$}
&\multicolumn{1}{c}{$10$} \\
\hline
$10$ & $1\,000$ & $977 $ & $816 $ & $627$ & $460$ & $358 $ & $271 $ & $222 $ & $158 $ & $125$ \\
$20$ & $1\,000$ & $1\,000$ & $1\,000$ & $985$ & $929$ & $830 $ & $720 $ & $618 $ & $524 $ & $450$ \\
$30$ & $178 $ & $1\,000$ & $1\,000$ & $1\,000$ & $999$ & $988 $ & $958 $ & $894 $ & $826 $ & $761$ \\
$40$ & $0 $ & $999 $ & $1\,000$ & $1\,000$ & $1\,000$ & $1\,000$ & $998 $ & $985 $ & $962 $ & $933$ \\
$50$ & $0 $ & $733 $ & $1\,000$ & $1\,000$ & $1\,000$ & $1\,000$ & $1\,000$ & $999 $ & $994 $ & $983$ \\
$60$ & $0 $ & $1 $ & $1\,000$ & $1\,000$ & $1\,000$ & $1\,000$ & $1\,000$ & $1\,000$ & $998$ & $996$ \\
$70$ & $0 $ & $0 $ & $879 $ & $999$ & $1\,000$ & $1\,000$ & $1\,000$ & $1\,000$ & $1\,000$ & $1\,000$ \\
$80$ & $0 $ & $0 $ & $0 $ & $927$ & $993$ & $1\,000$ & $1\,000$ & $1\,000$ & $1\,000$ & $1\,000$ \\
$90$ & $0 $ & $0 $ & $0 $ & $537$ & $859$ & $983$ & $998$ & $1\,000$ & $1\,000$ & $1\,000$ \\
$100$ & $0 $ & $0 $ & $0 $ & $72$ & $405$ & $818 $ & $946$ & $988 $ & $998 $ & $1\,000$ \\
\hline
\end{tabular}

\end{table}

In Table~\ref{table1}, a ``main diagonal'' is discernible whose entries all
equal $1\,000$ and thus indicate combinations of $\vert N\vert$ and $d$ values
for which none of the graph instances failed. Above this diagonal failures occur
because $L$ turns out empty, which occurs more frequently as $\vert N\vert$ is
decreased or $d$ is increased. Failures below the diagonal occur for at least
one of the remaining two reasons, viz., an excessive number of edges in $L$ or
an excessive number of feasible matchings in $\mathcal{M}$. Both forms of
failure become more frequent with increasing $\vert N\vert$ or decreasing $d$,
but failure by too many edges in $L$ is by far the most frequent of the two.

Given a $\vert N\vert,d$ pair, enumerating all the feasible matchings in
$\mathcal{M}$ for a passing graph instance $G$ has taken on average up to about
$597$ seconds to complete. Running Step~1 to find $\chi'^*_\mathrm{phys}(G)$, or
Step~3 to find $\chi'_\mathrm{phys}(G)$ whenever reaching that step, has
required on average up to about $199$ and $3\,575$ seconds, respectively. These
figures refer to an Intel Xeon E5-1650 v4 running at 3.6 GHz on 128 GB of RAM.
Such maximum averages were all observed for $\vert N\vert=100$ and $d=4$ km. In
view of these running times, it is unlikely that problem instances can be scaled
up significantly while still being amenable to solution by exact methods. We
return to this issue in Section~\ref{scale}.

\subsection{Examples}
\label{examples}

\begin{figure}[p]
\centering
\includegraphics[scale=1.10]{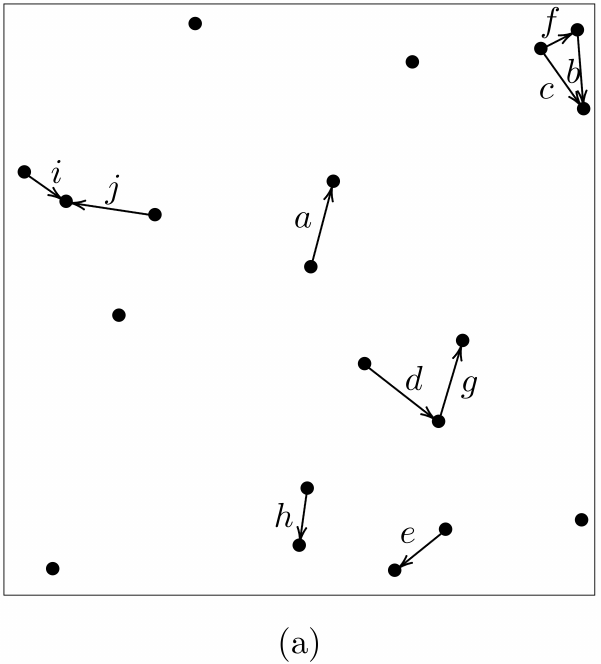}\\
\vspace{0.25in}
\includegraphics[scale=1.10]{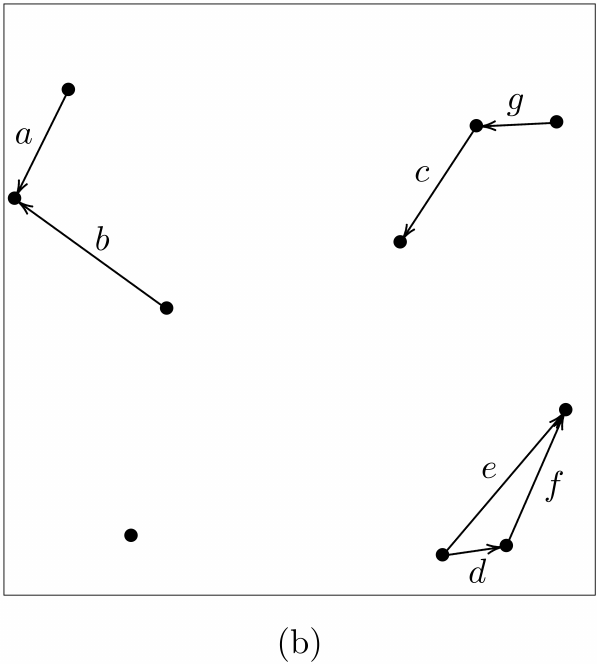}
\caption{Sample graph instances $G$. The one in panel (a) was generated for
$\vert N\vert=20$ and $d=2$ km, and turned out to have $\vert L\vert=10$ and
$\vert\mathcal{M}\vert=29$. Its edge-chromatic properties were found to be such
that $\chi'^*_\mathrm{phys}(G)=\chi'_\mathrm{phys}(G)=6$. The graph instance in
panel (b) was generated for $\vert N\vert=10$ and $d=1$ km, and has
$\vert L\vert=7$ and $\vert\mathcal{M}\vert=10$. Its edge-chromatic properties
are such that $\chi'^*_\mathrm{phys}(G)=11/2<6=\chi'_\mathrm{phys}(G)$. Edges
are drawn with directions to highlight their incident vertices' roles as either
sender nodes (the edges' tail vertices) or receiver nodes (the edges' head
vertices).}
\label{samplenets}
\end{figure}

Having described our methods to generate graph instance $G$ and to solve the
corresponding LP and ILP problems, it is worth returning to the discussion of
Section~\ref{steps}, with examples aiming to clarify the relationship between
$\chi'^*_\mathrm{phys}(G)$ and $\chi'_\mathrm{phys}(G)$ vis-\`a-vis the values
of those problems' variables at the optima they report. Two examples are given
in Figure~\ref{samplenets}. They both contain non-unit variables in the solution
to the LP problem, but $\chi'^*_\mathrm{phys}(G)$ relates differently to
$\chi'_\mathrm{phys}(G)$ in each case.

The first example, given in panel (a) of the figure, refers to a graph instance
$G$ for which $\chi'^*_\mathrm{phys}(G)=\chi'_\mathrm{phys}(G)=6$. Detecting
this, however, required solving both the LP problem (to discover the value of
$\chi'^*_\mathrm{phys}(G)$) and the ILP problem (to discover the value of
$\chi'_\mathrm{phys}(G)$). Solving only the former problem and inspecting its
variables' values at the optimum revealed eight feasible matchings $M$ for which
$x_M\neq 0$, four of them with $x_M=1$ (matchings $\{a\}$, $\{d\}$, $\{g\}$, and
$\{b,j\}$), four others with $x_M=1/2$ (matchings $\{e,f\}$, $\{c,h\}$,
$\{c,e,i\}$, and $\{f,h,i\}$). This illustrates why looking for non-unit
variables at the optimum as a proxy for
$\chi'^*_\mathrm{phys}(G)<\chi'_\mathrm{phys}(G)$ can be misleading. In fact, in
the case in question, solving the ILP problem brought up the possibility of
ending up with six nonzero variables, all equal to $1$, corresponding to
matchings $\{a\}$, $\{d\}$, $\{c,e\}$, $\{g\}$, $\{f,h,i\}$, and $\{b,j\}$. Had
this been the solution returned to the LP problem to begin with, we would have
known that $\chi'^*_\mathrm{phys}(G)=\chi'_\mathrm{phys}(G)$ immediately and
would have been able to dispense with the need to solve the ILP problem as well.

The example in panel (b), on the other hand, is such that
$\chi'^*_\mathrm{phys}(G)=11/2<6=\chi'_\mathrm{phys}(G)$, serving to illustrate
those cases in which detecting non-unit variables at the optimum of the LP
problem does indeed translate into a situation of
$\chi'^*_\mathrm{phys}(G)<\chi'_\mathrm{phys}(G)$ (even though, as in the
previous example, this can only be known after the ILP problem is solved as
well). In the case of Figure~\ref{samplenets}(b), the solution to the LP problem
indicated seven feasible matchings $M$ for which $x_M\neq0$, four with $x_M=1$
(matchings $\{b\}$, $\{c\}$, $\{e\}$, and $\{f\}$), three with $x_M=1/2$
(matchings $\{a,d\}$, $\{a,g\}$, $\{d,g\}$). As for the solution to the ILP
problem, six nonzero variables were identified at the end, all equal to $1$,
corresponding to matchings $\{b\}$, $\{c\}$, $\{d\}$, $\{e\}$, $\{f\}$, and
$\{a,g\}$.

\begin{figure}[t]
\centering
\includegraphics[scale=1.10]{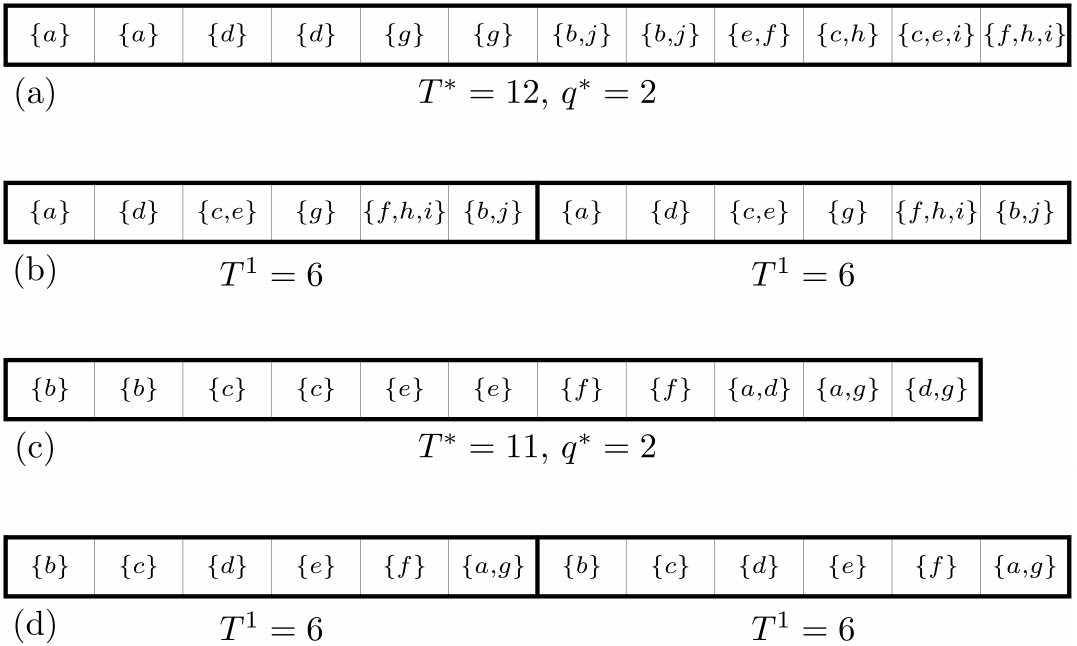}\\
\caption{Schedules corresponding to the optimal solutions to the LP and ILP
problems for the two examples of Figure~\ref{samplenets}. Panels (a) and (b)
correspond to the example of Figure~\ref{samplenets}(a), the former panel
depicting a single occurrence of the $T^*$-slot schedule implied by the solution
to the LP problem, the latter depicting $q^*$ repetitions of the $T^1$-slot
schedule implied by the solution to the ILP problem. We have $T^*=T^1q^*$, so
the two are essentially equivalent. Panels (c) and (d) refer, correspondingly,
to the example of Figure~\ref{samplenets}(b). Now we have $T^*<T^1q^*$, so the
$T^*$-slot schedule is preferable.}
\label{scheds}
\end{figure}

Further insight can be gained into the examples of Figure~\ref{samplenets} by
considering the actual schedules implied by the solutions to the LP and ILP
problems. In either example, solving the corresponding LP problem yielded either
$x_M=1$ or $x_M=1/2$ for all nonzero variables at the optimum. As discussed in
Section~\ref{LP}, this implies that every matching $M$ for which $x_M=1$ is to
appear in the resulting schedule with multiplicity $T_M=2$, while those with
$x_M=1/2$ appear with multiplicity $T_M=1$. The number of slots in the schedule
is the sum $T^*$ of all multiplicities. The number of times each edge appears in
these $T^*$ slots, denoted by $q^*$, is the least common multiple of the
denominators of all nonzero $x_M$'s at the optimum (in either example at hand,
$q^*=2$). As for the solution to each ILP problem, the schedule it implies has a
number $T^1$ of slots given by the sum of all $x_M$'s at the optimum. It is
better to use multiple colors per edge whenever $T^*<T^1q^*$. This is not the
case of the example in Figure~\ref{samplenets}(a), whose schedules are shown in
Figures~\ref{scheds}(a) and~(b), but is the case of the example in
Figure~\ref{samplenets}(b), whose schedules appear in
Figures~\ref{scheds}(c) and~(d).

\section{Results and discussion}
\label{results}

\begin{figure}[t]
\centering
\includegraphics[scale=0.95]{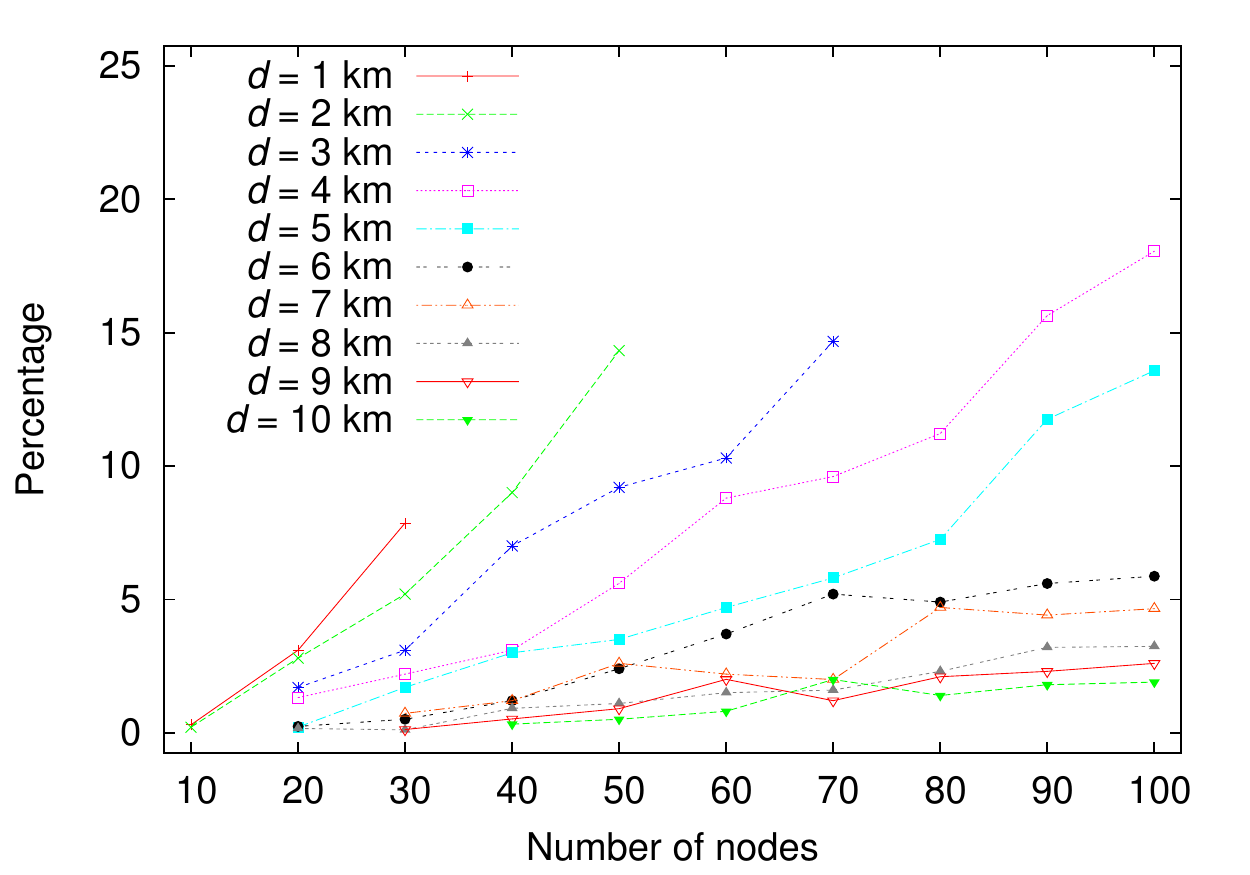}
\caption{Percentage of the graph instances $G$ in Table~\ref{table1} for which
$\chi'^*_\mathrm{phys}(G)<\chi'_\mathrm{phys}(G)$. Combinations of
$\vert N\vert$ and $d$ values for which the table reports a positive number but
which yield no such instances are omitted from the corresponding plots. This is
the case of all points missing for $\vert N\vert=10$, $20$, or $30$, as well as
for $\vert N\vert=60$ with $d=2$ km.}
\label{howmany}
\end{figure}

\begin{figure}[t]
\centering
\includegraphics[scale=0.95]{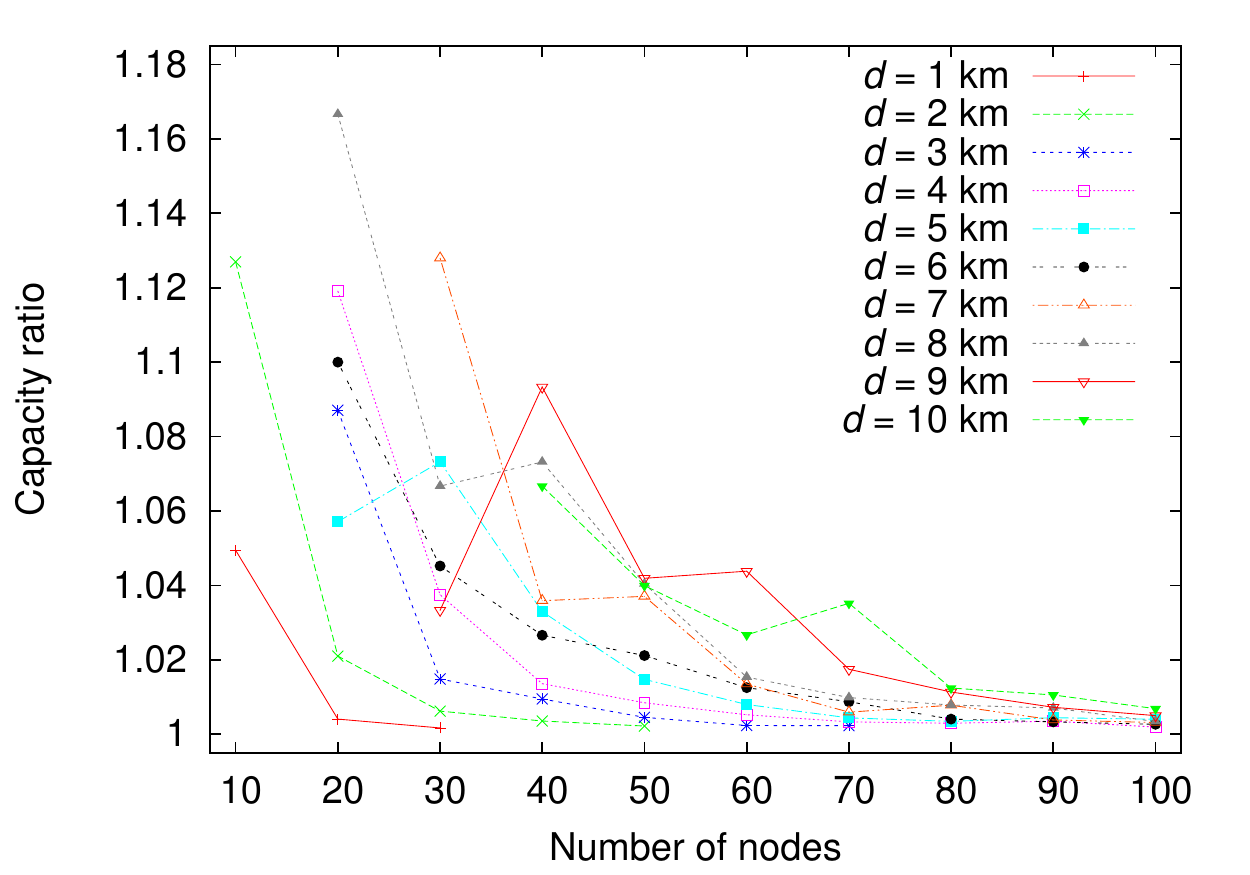}
\caption{Average value of the capacity ratio
$\chi'_\mathrm{phys}(G)/\chi'^*_\mathrm{phys}(G)$ over the graph instances $G$
in Table~\ref{table1} for which
$\chi'^*_\mathrm{phys}(G)<\chi'_\mathrm{phys}(G)$ (i.e., those accounted for in
Figure~\ref{howmany}). Confidence intervals are shown separately in
Figure~\ref{confint}, for the sake of clarity.}
\label{howmuch}
\end{figure}

\begin{figure}[t]
\centering
\includegraphics[scale=0.95]{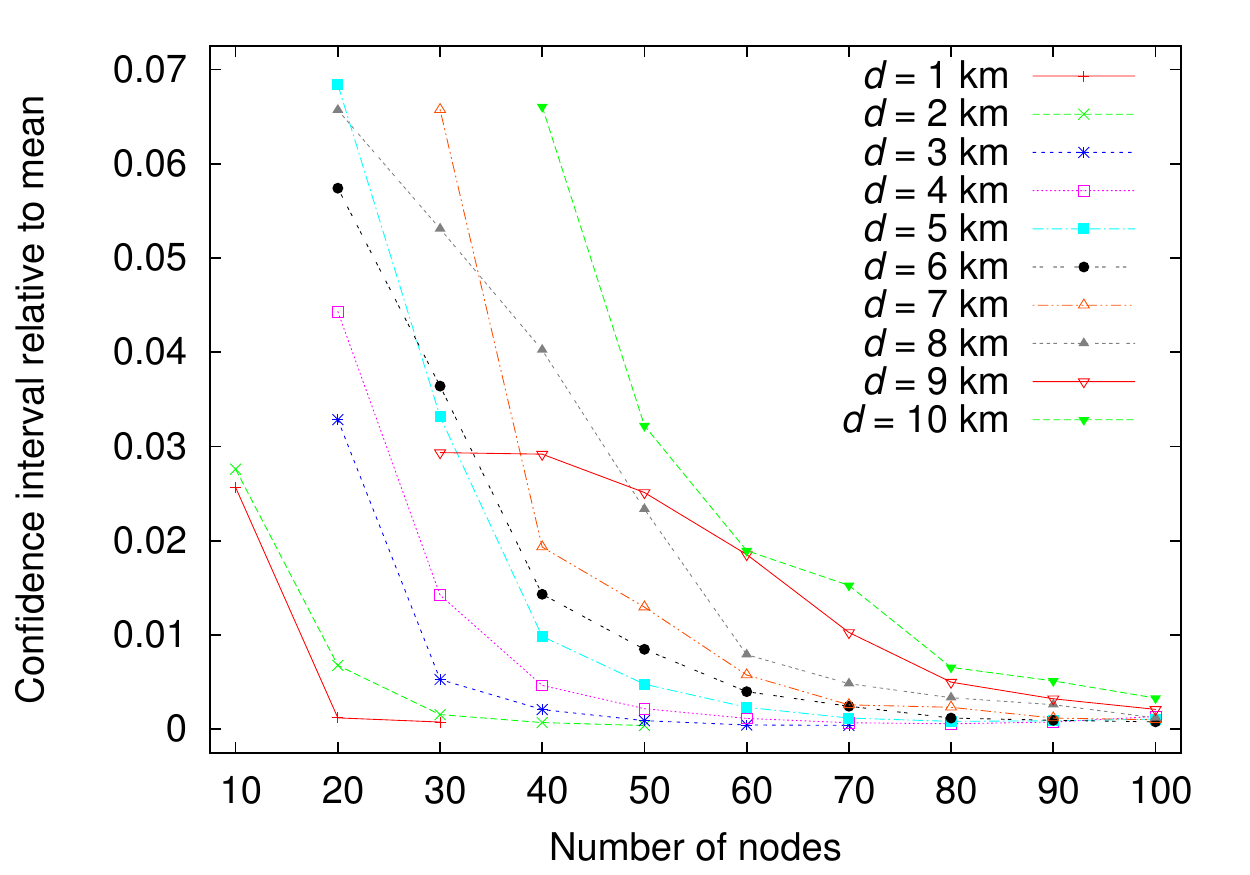}
\caption{Confidence intervals corresponding to the data in Figure~\ref{howmuch}.
Intervals are given at the $95\%$ level as fractions of the corresponding
means.}
\label{confint}
\end{figure}

We give results in Figures~\ref{howmany},~\ref{howmuch}, and~\ref{confint}, with
Figure~\ref{howmany} showing the percentage of those graph instances $G$ given
in Table~\ref{table1} for which
$\chi'^*_\mathrm{phys}(G)<\chi'_\mathrm{phys}(G)$. Such graph instances are
cases of network-capacity improvement when substituting a schedule based on
fractional edge coloring (of capacity $1/\chi'^*_\mathrm{phys}(G)$) for one
based on edge coloring that employs one single color per edge (of capacity
$1/\chi'_\mathrm{phys}(G)$). The ratios of capacity improvement, given by
$\chi'_\mathrm{phys}(G)/\chi'^*_\mathrm{phys}(G)$, are shown in
Figure~\ref{howmuch} as averages over the pertinent graph instances $G$. The
corresponding confidence intervals are given in Figure~\ref{confint} as
fractions of the corresponding means.

If we define a geometric graph's node density to be its number of nodes divided
by the area of deployment, then clearly a tendency is shown in
Figure~\ref{howmany} of higher percentages for higher node densities. This is
easily seen as we fix $d$ while $\vert N\vert$ is increased, but holds across
values of $d$ as well: note that node density is of the order of $10^{-5}$ to
$10^{-4}$ for $d=1$ km, $10^{-6}$ to $10^{-5}$ for $d=2$ km, and so on, which in
general correlates well with lower-$d$ plots being located higher up in the
figure. Of course, owing to the total absence of graph instances for the highest
values of $\vert N\vert$ and lowest values of $d$ in Table~\ref{table1}, at this
point we can only speculate as to what would happen if such instances' number of
edges and of feasible matchings could be handled, but the trend seems clear
nonetheless. Indeed, increasing a geometric graph $G$'s node density tends to
lead to a higher number of edges, and therefore a pressure exists for the value
of $\chi_\mathrm{phys}'(G)$ to increase as well. Intuitively, this presents an
opportunity for some $k>1$ to prevail in Eq.~(\ref{chi*phys}) and for
$\chi_\mathrm{phys}'^*(G)<\chi_\mathrm{phys}'(G)$ to occur.

Another observable of interest is the ratio of capacity improvement, given by
$\chi_\mathrm{phys}'(G)/\chi_\mathrm{phys}'^*(G)$, for those instances $G$ for
which $\chi_\mathrm{phys}'^*(G)<\chi_\mathrm{phys}'(G)$ is obtained. This is
shown in Figure~\ref{howmuch} as averages over the instances accounted for in
Figure~\ref{howmany}, with relative confidence intervals shown separately in
Figure~\ref{confint} (for clarity's sake). A relationship continues to exist
between the graphs' node densities and their capacity gains, but unlike the case
of Figure~\ref{howmuch}, now the trend is for the lower-node-density graphs to
afford higher capacity gains. This can be seen as we fix $d$ and increase
$\vert N\vert$ (plots in the figure are generally decreasing toward $1$) and, to
a limited extent, across values of $d$ as well. What prevents us from stating
the latter more firmly is the way the plots in Figure~\ref{howmuch} deviate from
what they would look like ideally (plots nicely nested one above the other with
increasing $d$). This may have to do with the higher confidence intervals
occurring precisely where deviations from the said ideal are most striking
(confidence intervals up to nearly $7\%$ of the corresponding means in some
cases; cf.\ Figure~\ref{confint}), but only further experimentation will clarify
the issue.

\section{Concluding remarks}
\label{concl}

In this paper we have addressed the problem of maximizing the capacity of
wireless mesh networks by scheduling its links optimally for operation in the
physical interference model. We have modeled the network as an undirected graph
$G$ and cast the scheduling problem as the problem of coloring the edges of $G$.

In the simpler, protocol-based interference model, the customary approach is to
define network capacity as $1/\chi'(G)$, where $\chi'(G)$ is the chromatic index
of $G$ (i.e., the minimum number of colors with which $G$'s edges can be colored
with one color per edge, provided no two edges sharing an end vertex get the
same color). By redefining network capacity as $1/\chi'^*(G)$, where
$\chi'^*(G)$ is the fractional chromatic index of $G$, and noting that
$\chi'^*(G)\le\chi'(G)$ necessarily, it becomes possible to aim for higher
network capacity as links are scheduled for operation. Fractional coloring
differs from integer coloring in that edges are allowed to receive more than
one color, the same number of colors for all edges, and also in that optimality
is now defined as minimizing the ratio $\chi'^k(G)/k$ for $k\ge 1$, where
$\chi'^k(G)$ is the minimum number of colors required to color the edges of $G$
in such a way that every edge gets $k$ colors.

Finding $\chi'^*(G)$ can be approached via the solution of an LP problem based
on knowing the set $\mathcal{M}$ of all matchings of graph $G$. This makes the
transition to the physical interference model straightforward by simply letting
$\mathcal{M}$ contain only those matchings that are feasible as defined for the
model. Concentrating on this restricted set of matchings led to the definition
of a new fractional chromatic index for $G$, $\chi'^*_\mathrm{phys}(G)$, and
correspondingly a new definition of network capacity,
$1/\chi'^*_\mathrm{phys}(G)$. Notably, the single-color-per-edge definitions of
network capacity, $1/\chi'(G)$ for the protocol-based interference model,
$1/\chi'_\mathrm{phys}(G)$ for the physical interference model, are not
eliminated by the new definitions. Instead, they are simply subsumed, because
they correspond to the $k=1$ cases and can therefore be optimal whenever
$\chi'^*(G)=\chi'(G)$ or $\chi'^*_\mathrm{phys}(G)=\chi'_\mathrm{phys}(G)$,
respectively.

Our computational experiments on the physical interference model do confirm that
this can happen relatively often. On the other hand, they also reveal that,
particularly in the case of denser networks (larger number of nodes per unit
area), occurrences of $\chi'^*_\mathrm{phys}(G)<\chi'_\mathrm{phys}(G)$ happen
as well and sometimes account for non-negligible capacity improvements (i.e.,
increases in the ratio $\chi'_\mathrm{phys}(G)/\chi'^*_\mathrm{phys}(G)$). Be
that as it may, adopting the fractional-coloring framework allows optimization
to be carried out exclusively by solving the corresponding LP problem, without
any need whatsoever to call upon the ILP problem that accompanies it. As we
remarked in Section~\ref{setup}, solving the latter is in general substantially
more time-consuming.

We close with further remarks on issues that were left open in the previous
sections.

\subsection{Multigraph representation of the network}
\label{multi}

As we remarked in Section~\ref{math}, representing the network by the undirected
graph $G$ precludes any two nodes from participating together in more than one
link, even in two ``antiparallel'' links (i.e., two links $e,f$ in which one of
the nodes serves as sender in $e$ and receiver in $f$, the other node as sender
in $f$ and receiver in $e$). This representational difficulty can be easily
resolved by resorting to a multigraph instead of a graph, i.e., by allowing two
vertices to be joined by multiple edges.

The consequence of this for our formulation in Section~\ref{math} would be
simply to increase the size of the matching set $\mathcal{M}$. In fact, if we
denote this set by $\mathcal{M}_1$ when all edges have unit multiplicities and
by $\mathcal{M}_\mathrm{mult}$ when a general multiplicity $m_e$ is allowed for
each edge $e\in L$, then the effect would be an increase in the overall number
of matchings from $\vert{\mathcal{M}_1}\vert=\sum_{M\in\mathcal{M}_1}1$ to
$\vert\mathcal{M}_\mathrm{mult}\vert=\sum_{M\in\mathcal{M}_1}\prod_{e\in M}m_e$.

\subsection{Regarding scalability}
\label{scale}

It is clear from our discussion in Section~\ref{results} that the networks we
experimented with were limited by the need to enumerate all feasible matchings
in $\mathcal{M}$ in order to exactly solve the LP problem given by
Eqs.~(\ref{xobj})--(\ref{xsumis1}) and the ILP problem given by
Eqs.~(\ref{xobjint})--(\ref{xsumis1int}). The number of such matchings
eventually exhausts all computational resources available, thus making it
impossible for $\mathcal{M}$ to be enumerated and for either problem to be
solved. Ultimately, however, it is the LP problem that matters most, and for
this one a clear path exists for the search for scalability.

To see that a substantially more efficient alternative may be available,
consider the particular case mentioned at the beginning of
Section~\ref{ecoloring}, in which $\mathcal{M}$ is the set of all matchings of
$G$ (i.e., not necessarily feasible in the sense of the physical interference
model). In this case, the better alternative is to consider the LP problem in
its dual formulation, given as follows.
\begin{align}
\text{maximize } &\textstyle z=\sum_{e\in L}y_e & \label{yobj}\\
\text{subject to } &\textstyle\sum_{e\in M}y_e\le 1. &\forall M\in\mathcal{M}
\label{ysumisle1}
\end{align}
In this formulation, for each $e\in L$ we have a real variable $y_e$ (which can
be negative, zero, or positive, by virtue of the equality constraint in
Eq.~(\ref{xsumis1})), and for each matching $M\in\mathcal{M}$ we have a
constraint forbidding the $y_e$'s for $e\in M$ to add up to more than $1$
(Eq.~(\ref{ysumisle1})). The goal is to maximize the sum of all $y_e$'s (the
objective function $z$ in Eq.~(\ref{yobj})). By LP duality, if the LP problem in
Eqs.~(\ref{xobj})--(\ref{xsumis1}) defines the fractional chromatic index of
$G$, then so does the one in Eqs.~(\ref{yobj}) and~(\ref{ysumisle1}).

It would seem that the new formulation suffers from the same problem as the
previous one, the only difference being that now the size of $\mathcal{M}$ is
reflected in the number of constraints, not the number of variables. While the
latter is clearly true, it is in principle possible to solve the problem without
listing all constraints explicitly. We start by maximizing $z$ subject to only a
minimal set of constraints (one for each singleton matching $\{e\}\subseteq L$,
which ensures a finite maximum value for $z$). Then we iterate, each time
expanding the set of explicitly listed constraints with the addition of an
unlisted one that is currently violated. We do this until no violated unlisted
constraints remain.

In order to succeed with this approach, we must ensure that both the time
required to identify a violated unlisted constraint and the overall number of
iterations are polynomially bounded. The first of these goals is achieved by
resorting to the problem of finding a maximum-weight matching in $G$, which is
known to be solvable in polynomial time by a variety of methods (cf., e.g.,
\cite{hk17} and references therein). To see how this problem can be of use, let
$y_e^*$ be the value of $y_e$ for each $e\in L$ after one of the iterations and
find a maximum-weight matching of $G$ with the $y_e^*$'s as weights. Let $M^*$
be the matching obtained, of weight $W^*=\sum_{e\in M^*}y_e^*$. If $W^*>1$, then
clearly the constraint in Eq.~(\ref{ysumisle1}) for $M=M^*$ is being violated
and should therefore be listed explicitly for the next iteration. If $W^*\le 1$,
then clearly no further violated constraints exist (since $M^*$ has maximum
weight) and no further iterations are needed. As for the second goal, that of
iterating for only a polynomially-bounded number of times, the ellipsoid method
for linear programming, though impractical, provides the necessary guarantee
\cite{gls81}. An essentially equivalent path is followed in \cite{hs88}.

The case in which the matchings in $\mathcal{M}$ are all feasible in the sense
of the physical interference model is substantially more complex, but at least
we have the results of \cite{gls81} to rely on for guidance. Specifically, what
we must do is discover a polynomial-time algorithm to find a maximum-weight
feasible matching of $G$. Such an algorithm will depend on all the intricacies
underlying the definition of SINR in Eq.~(\ref{sinr}), and whether one exists is
for now an open problem. Should it not exist, or should it prove too elusive to
find, a more costly algorithm will also do: though requiring more computational
effort to determine the required feasible matching, the expected savings derived
from not having to list a huge number of constraints explicitly are bound to be
worth the additional resources it expends.

\subsection*{Acknowledgments}

We acknowledge partial support from Conselho Nacional de Desenvolvimento
Cient\'\i fico e Tecnol\'ogico (CNPq), Coordena\c c\~ao de Aperfei\c coamento de
Pessoal de N\'\i vel Superior (CAPES), and a BBP grant from Funda\c c\~ao Carlos
Chagas Filho de Amparo \`a Pesquisa do Estado do Rio de Janeiro (FAPERJ).

\bibliography{fracsched}
\bibliographystyle{unsrt}

\end{document}